\documentclass[a4paper]{jpconf}
\usepackage{graphicx}
\usepackage{hyperref}

\newcommand{\be}{\begin{equation}}
\newcommand{\ee}{\end{equation} }
\newcommand{\beqa}{\begin{eqnarray} }
\newcommand{\eeqa}{\end{eqnarray} }
\newcommand{\ba}{\begin{array}}
\newcommand{\ea}{\end{array}}
\newcommand{\bpm}{\begin{pmatrix}}
\newcommand{\epm}{\end{pmatrix}}

\newcommand\Tp{{\cal{T}}_{P}}
\newcommand\Tv{{\cal{T}}_{V}}
\newcommand\Vp{{\cal{V}}_{P}}
\newcommand\BEC{{\rm\scriptscriptstyle{BEC}}}
\newcommand\kB{k_{{\scriptscriptstyle{\rm B}}}}

\begin{document}
\title{How many is different?  Answer from ideal Bose gas}

\author{Jeong-Hyuck Park}

\address{Department of Applied Mathematics and Theoretical Physics, Cambridge, CB3 0WA, England\\
Department of Physics, Sogang University,  Seoul 121-742, Korea\footnote{Sabbatical leave of absence.}}

\ead{park@sogang.ac.kr}

\begin{abstract}
How many $\mathrm{H_{2}O}$ molecules are needed to form water? While the precise answer is not known, it is clear that the answer should be a finite number rather than infinity.  We   revisit with care  the ideal Bose gas confined in a cubic box which is discussed in most statistical physics textbooks.   We show that the isobar of the ideal gas zigzags on the temperature-volume  plane  featuring a \textit{boiling-like} discrete phase transition,    provided the number of particles is equal to or greater than a particular value:   $7616$. This  demonstrates for the first time    how  a finite system can feature a mathematical singularity  and  realize  the notion of `Emergence',  without resorting to  the thermodynamic limit.
\end{abstract}


\section{Introduction}
Emergence is a generic notion that quantitative increase leads to qualitative change. It is often said that the whole is greater than the sum of its parts. It is a key idea in condensed matter physics as well as in statistical physics, such that a classic paper by Anderson goes with the title, \textit{More is different}~\cite{Anderson}. Then  the question we address in this note is \textit{How many is different?} To answer the question, we consult the  quantum statistical physics  where the key quantity is a partition function,
\be
Z=\mathrm{Tr}\left[e^{- H/(\kB T)}\right]\,.
\ee
The energy eigenvalues of the Hamiltonian are generically quantized with respect to the  size of the  volume. Consequently  the partition function depends on both the temperature, $T$, and the volume, $V$. 
Once we know the exact expression  of the partition  function we may compute all the physical quantities which  are typically given by the derivatives of the logarithm of the partition function.

For a finite system, the partition function and the derivatives are all analytic and hence all finite, never diverge. For example, $C_{V}$, the specific heat at constant volume is finite: no singularity arises.  The conventional  way to realize a singularity is then to assume the thermodynamic limit where the volume and the  number of particles  go to infinity, $V\rightarrow\infty$, $N\rightarrow\infty$, while the density, $N/V$, is kept fixed.  In this limit, the partition function may become non-analytic and feature a singular behaviour~\cite{YangLee}.   It might seem  that only infinity system could  feature a singularity and realize a discrete phase transition,  to coin a phrase,    ``More is the same: infinitely more is different"~\cite{Kadanoff}.

However, strictly speaking, infinite limits are hardly realistic and exist only in theory~\cite{Felderhof,Gross:2002mg,Butterfield}.   Experiments  or  our daily experiences,  like   the boiling water in the  kitchen, do appear  not just to mimic but actually to manifest    discrete phase transitions.  As a matter of historical fact in 1937, at the Van der Waals memorial meeting the audience could not agree on the question, whether partition function for a finite system could  explain a sharp phase transition or not. So the chairman of the session, Kramers, put it to a vote!\,\footnote{The author is not aware of the result of this  `democratic' decision.}

Nevertheless, despite  the old controversy,  in a modern rather axiomatic approach  adopted mainly  by mathematicians,   phase transitions are   defined  simply  in the thermodynamic limit only,  with the    ``mathematically-natural" conventional  answer    \textit{infinity}  to the question, \textit{How many is different?}   The present note highlights an   alternative  finite answer,  like $N=7616$ reported  in    the works of the author with Sang-Woo Kim and Imtak Jeon~\cite{7616,rel7616,14393}.\footnote{A short  video presentation  is  also available \href{http://www.youtube.com/watch?v=rNmOvgQsGnU&list=FLfVQVNKya2JpncnEth00MjA&feature=mh_lolz}{on-line}: \\
\url{http://www.youtube.com/watch?v=rNmOvgQsGnU&list=FLfVQVNKya2JpncnEth00MjA&feature=mh_lolz}}  The emphasis is put on physical implications rather than technical derivations.


\section{Discrete phase transition of a finite system   under constant  pressure}
A crucial  ingredient   in our approach to realize  a discrete  phase transition from  a finite system is to \textit{keep not the volume but the pressure constant}.  
The temperature derivative at fixed pressure acting on an arbitrary  function of $T$ and $V$ reads
\be
\left.\frac{\partial~}{\partial T}\right|_{P}=
\left.\frac{\partial~}{\partial T}\right|_{V}-\left[\frac{\partial_{T}P(T,V)}{\partial_{V}P(T,V)}\right]\left.\frac{\partial~}{\partial V}\right|_{T}\,\,.
\ee
When this differential operator acts on  the logarithm of the analytic partition function of a finite system,  the temperature and the volume derivatives  on the right hand side of the equality cannot generate any mathematical singularity. However, when $\partial_{V}P(T,V)$ vanishes, namely at the `spinodal' points,  it can be singular.  In fact, the usual argument against the mathematical singularity from  a finite system assumes the volume to be fixed. Alternatively,  if we fix the pressure constant, a singularity may occur. That is to say, for a finite system, $C_{V}$ is finite never  diverges, but $C_{P}$ may become singular,
\be
\ba{lllll}
C_{V}\ll\infty\quad&\mbox{versus}&\quad C_{P}=\infty\quad&\mbox{for}&\quad N\ll\infty\,.
\ea
\ee

Physically, if we fully fill a rigid box with water  and heat it up, the temperature will increase but the water hardly evaporates: no discontinuous phase transition to occur in this case.  Nevertheless, opening the lid will set the pressure  constant or at $1$  atm, and the water will surely start to boil at $100$ degree Celsius: no need to take the thermodynamic limit!

Our   main point    is that a singularity or a discrete phase transition may occur even for a finite system, provided  we impose the constraint of keeping the pressure constant, rather than  fixing  the volume or the density.  The next question is then the existence of the spinodal points where ${\partial_{V}P(T,V)=0}$.  Our  claim   is that the spinodal curve   originates essentially      due to  the \textit{identical nature of particles}  without need of any extra  interactions. As we describe below, a careful study of  an ideal Bose gas indeed reveals the spinodal points and hence a \textit{boiling-like} discrete  phase transition for finite $N$.


\section{Ideal Bose gas can boil under constant pressure if $N\geq7616$}
We are interested in a canonical partition function, $Z_{N}(T,V)$, of the ideal Bose gas confined in a cubic box with a definite number of particles, $N$.  It  is a  standard quantum  mechanical system discussed in most  of statistical physics textbooks, \textit{e.g.}~\cite{Huang}. We aim to   address the precise dependence on $N$.   The commonly adopted  approximation of the  partition function is
\be
Z_{N}\,\sim\,  \frac{\left(Z_{N=1}\right)^{N}}{N!}\,,
\ee
such that, \textit{e.g.}  the division by $N!$   solves  the Gibbs paradox. This approximation gives for a generic derivative,
\be
\partial\ln Z_{N}\,\sim\, N\partial\ln Z_{N=1}\,,
\ee
and hence the number  of particles   appears as an overall   parameter corresponding to an extensive quantity.    However, the above approximation would be only  correct   if the $N$ particles occupied always  $N$ distinct quantum states.  But if some of the  particles are in a common state, the devision by $N!$  is too much and  the approximation requires  improvement.  In fact,   the correction can be organized as a determinant of a certain $N\times N$ matrix, $\Omega_{N}$, and  the \textit{exact} formula  of the canonical partition function can be expressed as
\be
Z_{N}=\mathrm{Det}\left(\Omega_{N}\right)\times\frac{\left(Z_{N=1}\right)^{N}}{N!}\,.
\ee
For the details  we refer the readers  to \cite{7616,rel7616}. The point   is that with the exact expression,   $N$ is no longer a strictly-extensive parameter: physical quantities    depend \textit{non-linearly} on the number of particles.  Especially a  thermodynamic instability and hence  zigzagging  of the isobar on $(T,V)$-plane \textit{emerge}  provided the number of particles is equal to or greater than  $7616$~\cite{7616}.

\begin{figure}[h]
\begin{minipage}{16pc}
\includegraphics[width=15pc]{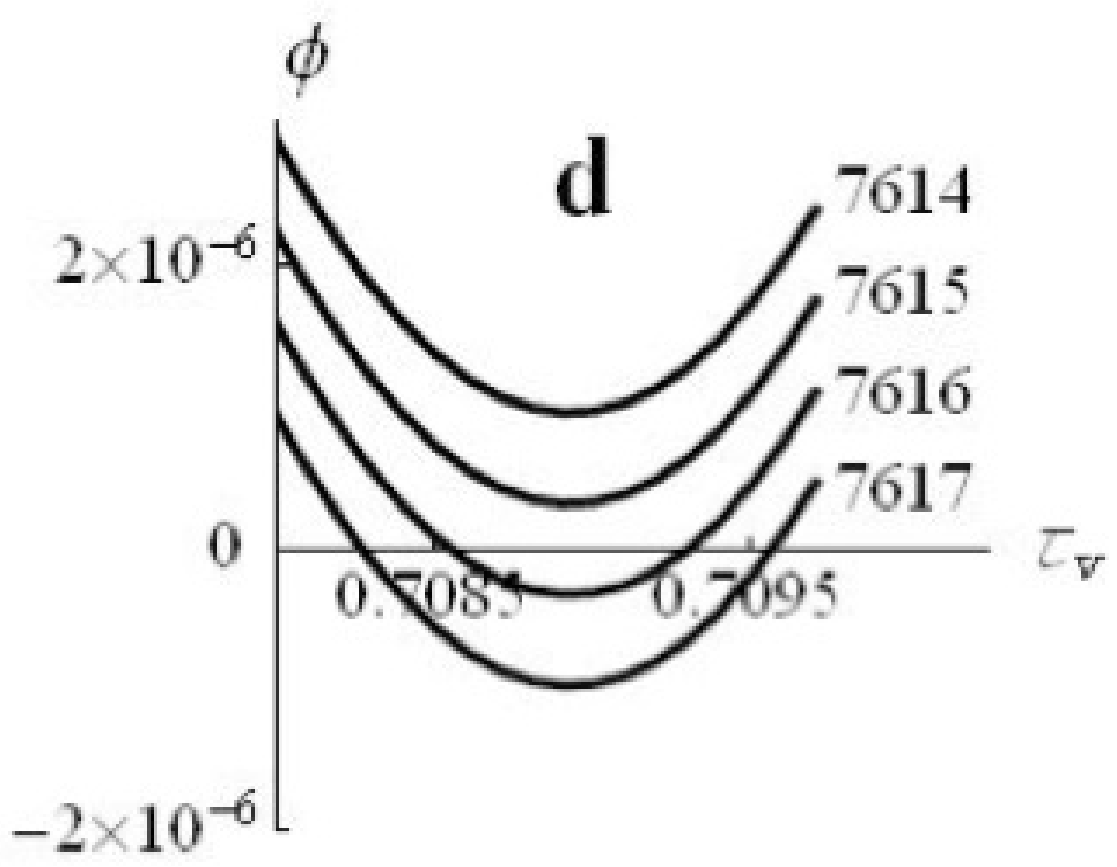}
\caption{\label{phi7616}A supercomputer-powered  numerical analysis  on  the exact   partition function reveals that a  thermodynamically unstable region  \textit{emerges} on $(T,V)$-plane  when $N\geq7616$.  }
\end{minipage}\hspace{1pc}%
\begin{minipage}{21pc}
\includegraphics[width=21pc]{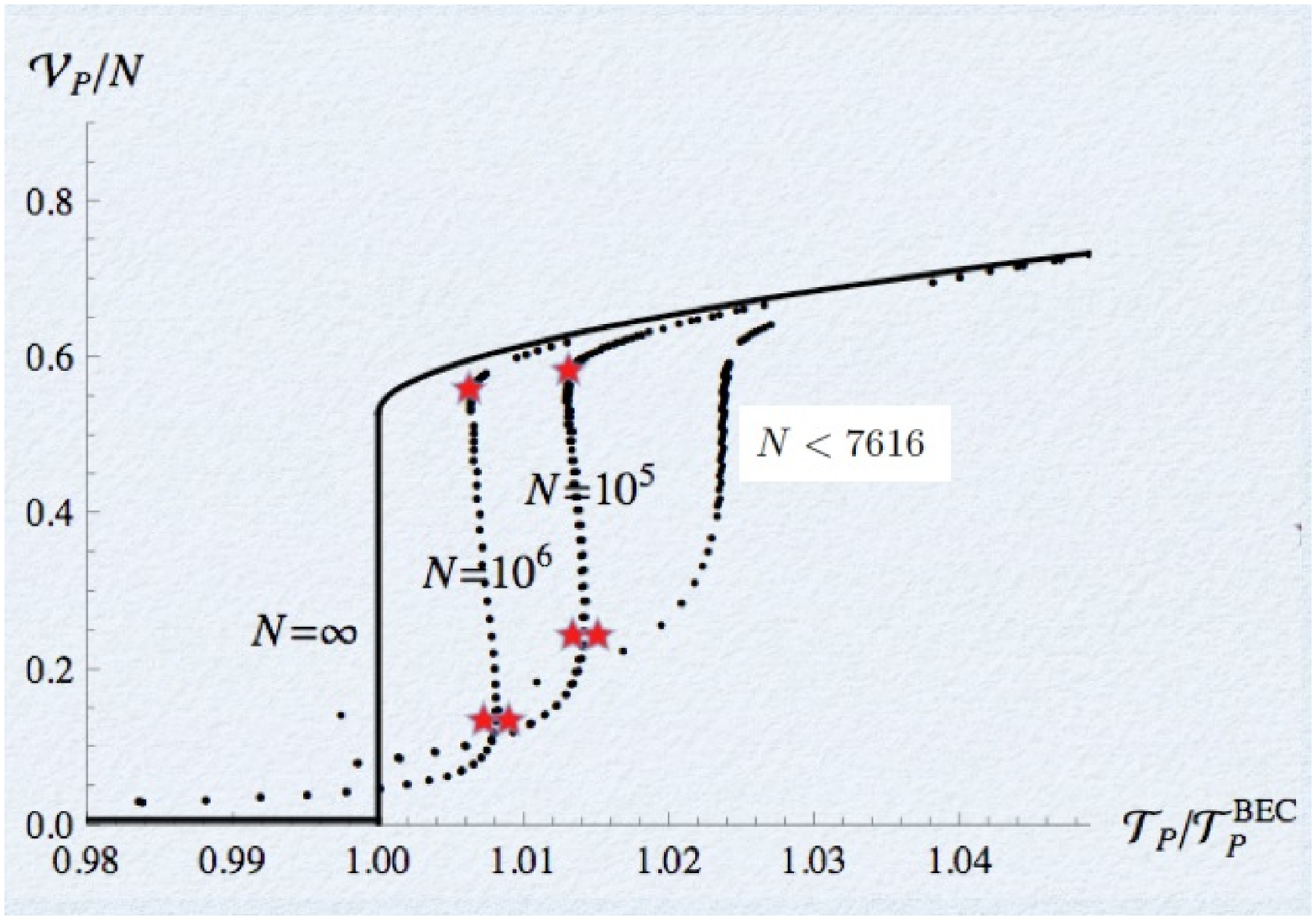}
\caption{\label{isobar7616}The isobar of the ideal Bose gas zigzags on $(T,V)$-plane when $N\geq7616$, which  features a \textit{boiling-like} discrete  phase transition. }
\end{minipage} 
\end{figure}
\noindent In the figures above, utilizing the scale symmetry of the system, we set  dimensionless quantities:
\be
\ba{llll}
\phi=-\left(\frac{V^2}{N\kB T}\right)\partial_{V}P\,,&\,
{\cal{T}}_{V}=\kB T\left(\frac{V}{N}\right)^{\frac{2}{3}}\!\left(\frac{2m}{\pi^{2}\hbar^{2}}\right)\,,&\,
{\cal{V}}_{P}=V\left(P\frac{2m}{\pi^{2}\hbar^{2}}\right)^{\frac{3}{5}}\,,&\,
{\cal{T}}_{P}=\kB TP^{-\frac{2}{5}}\left(\frac{2m}{\pi^{2}\hbar^{2}}\right)^{\frac{3}{5}}.
\ea
\ee
Further, for  $N\gg 7616$, the supercooling ($\ast$) and the superheating ($\ast\ast$) points are analytically~\cite{14393},
\be
\ba{ll}
\Tp^{\ast}/{\Tp^{\BEC}}\simeq 1+\textstyle{{\frac{\,\pi^{3}}{60}\left[\left(\Tp^{\BEC}\right)^{5}/\Tv^{\BEC}\right]^{\frac{1}{2}}}}\,N^{-\frac{1}{3}}\,,~&
\Tp^{\ast\ast}/{\Tp^{\BEC}}\simeq 1+\frac{1}{150}\left(\frac{\,\pi^{15}}{15}\right)^{\frac{1}{4}}
\left(\Tp^{\BEC}\right)^{\frac{5}{2}}\,N^{-\frac{1}{4}}\,,\\
\Vp^{\ast}\simeq \textstyle{\left({\Tv^{\BEC}}/{\Tp^{\BEC}}\right)^{\frac{3}{2}}\left(N+\frac{\pi}{4}\Tv^{\BEC}\,N^{\frac{2}{3}}\ln N\right)}\,,~&
\Vp^{\ast\ast}\simeq  8\left(\frac{15}{\,\pi^{3}}\right)^{\frac{3}{4}}
\left(\Tp^{\BEC}\right)^{-\frac{3}{2}}\,N^{\frac{3}{4}}\,,\\
\Tv^{\ast}\simeq {\Tv^{\BEC}}\left(1+\frac{\pi}{6}\Tv^{\BEC}\,N^{-\frac{1}{3}}\ln N\right)\,,~&
\Tv^{\ast\ast}\simeq 4\left(\frac{15}{\,\pi^{3}}\right)^{\frac{1}{2}}\,N^{-\frac{1}{6}}\,,
\ea
\ee
where  $\Tp^{\BEC}$ and $\Tv^{\BEC}$ denote  the ``BEC" constants,
\be
\ba{ll}
\textstyle{
\Tp^{\BEC}=\left(\frac{64}{\,\pi^{3}}\right)^{\frac{1}{5}}\left[\zeta(\frac{5}{2})\right]^{-\frac{2}{5}}}\simeq 1.02781\,,~&~
\textstyle{
\Tv^{\BEC}=\frac{4}{\pi}\left[\zeta(\frac{3}{2})\right]^{-\frac{2}{3}}\simeq 0.671253\,.}
\ea
\label{BECtemp}
\ee
\section{Discussion}
The above  boiling-like discrete phase transition  is an \textit{emergent phenomenon} of the finitely many bosonic identical particles, which we \textit{ab initio} derived from the first principles in statistical physics. The singularity is due to the spinodal curve that also sharply defines the phase diagram itself~\cite{rel7616},\textit{~c.f.~}\cite{Casetti:2000gd}.

Our results seem to suggest that, \textit{a generic liquid-gas phase transition occurs essentially due to the identical nature of particles} rather than additional  interactions.

The critical number $7616$ can be viewed as a quantum mechanically determined    characteristic  of `cube', the geometric shape of the box. Boxes of different shapes will have their own  critical numbers. For example,    for  `sphere'  it  is   $10458$.

It will be experimentally challenging to find a corresponding critical number for each molecule in Nature.  We believe   there must be a finite definite answer to the question, \textit{How many $\mathrm{H_{2}O}$ molecules are needed to form water that features boiling phenomenon with  zigzagged isobar?}

\ack{The author wishes to thank  Hugh Osborn at  DAMTP, University of Cambridge for      hospitality  and  encouragement    during  the  sabbatical visit.  This work was supported by the Sogang University Research Grant of 201310030.01.}

\end{document}